\documentclass{article}
\usepackage{spconf,amsmath,graphicx}

\usepackage{caption}
\usepackage{color}

\title{REDUCING MODEL COMPLEXITY FOR DNN BASED LARGE-SCALE AUDIO CLASSIFICATION}
\name{Yuzhong Wu, Tan Lee}
\address{Department of Electronic Engineering, The Chinese University of Hong Kong}
\begin{document}
%

%-----------------------------
% Upon publication of an article by the IEEE, the author must replace any previously posted electronic versions of the article with either (1) the full citation to the IEEE work with a Digital Object Identifier (DOI) or link to the article abstract in IEEE Xplore, or (2) the accepted version only (not the IEEE-published version), including the IEEE copyright notice and full citation, with a link to the final, published article in IEEE Xplore.

\newpage
\onecolumn

{\noindent \LARGE \textbf{IEEE Copyright Notice}} \\
~\\
{\noindent \Large $\copyright$ 2018 IEEE. \\
Personal use of this material is permitted. Permission from IEEE must be obtained for all other uses, in any current or future media, including reprinting/republishing this material for advertising or promotional purposes, creating new collective works, for resale or redistribution to servers or lists, or reuse of any copyrighted component of this work in other works. }

~\\
~\\

{\noindent \Large Accepted by 2018 IEEE International Conference on Acoustics, Speech and Signal Processing (ICASSP 2018), 15-20 April 2018, Calgary, Alberta, Canada}

~\\

{\noindent \Large Full citation: \\
Y. Wu and T. Lee, "Reducing Model Complexity for DNN Based Large-Scale Audio Classification," 2018 IEEE International Conference on Acoustics, Speech and Signal Processing (ICASSP), Calgary, AB, 2018, pp. 331-335. \\}

~\\

{\noindent \Large Link in IEEE Xplore: https://ieeexplore.ieee.org/document/8462168}

%-----------------------------

\twocolumn
\ninept
\maketitle
\begin{abstract}
Audio classification is the task of identifying the sound categories that are associated with a given audio signal. This paper presents an investigation on large-scale audio classification based on the recently released AudioSet database. AudioSet comprises $2$ millions of audio samples from YouTube, which are human-annotated with $527$ sound category labels. Audio classification experiments with the balanced training set and the evaluation set of AudioSet are carried out by applying different types of neural network models. The classification performance and the model complexity of these models are compared and analyzed. While the CNN models show better performance than MLP and RNN, its model complexity is relatively high and undesirable for practical use. We propose two different strategies that aim at constructing low-dimensional embedding feature extractors and hence reducing the number of model parameters. It is shown that the simplified CNN model has only $1/22$ model parameters of the original model, with only a slight degradation of performance.
\end{abstract}
\section{Introduction}
\label{sec:intro}

The rapid development of technology has made the production, rendering, sharing and transmission of multimedia data easy, low-cost and hence become part of our daily life. The growth of online available (public or restricted-access) audio and visual data is irreversible trend. Having effective and efficient tools for classifying, indexing and managing multimedia data is not only for the convenience and enjoyment of individuals, but also critical to the social and economic development in the big data era.

Audio is inarguably one of the most important types of multimedia resources to be reckoned. Audio classification is generally defined as the task of identifying a given audio signal from one of the predefined categories of sounds. Depending on the applications, the sound categories could be broad, e.g., music, voice, noise, or highly specified, e.g., children speech. The existence of diverse task definitions has made it difficult to compare the methods and results from different research groups and therefore hindered constructive exchange of ideas. In recent years, there have been organised efforts on setting up open evaluation or competitions on large-scale audio classification. For example, the IEEE AASP challenge DCASE2016 includes the acoustic scene classification (ASC) as an important part. The ASC task is to classify a $30$-second audio sample as one of the $15$ pre-defined acoustic scenes. Among the $30$ participating teams in DCASE2016 ASC, Eghbal-Zadeh et al. \cite{Eghbal-Zadeh2016} proposed a hybrid model using binaural I-vectors and CNN, and demonstrated the best performance. It was shown that log-mel filterbank features perform better than MFCC, when CNN models are applied \cite{Valenti2016}. In DCASE2017, CNN is most popular among the top-$10$ models. Mun et al. \cite{Mun2017} addressed the problem of data insufficiency and proposed to apply GAN-based data augmentation method to significantly improve the classification accuracy. There was also a trend on using binaural audio features rather than monaural features.

The DCASE2016 ASC task provides an annotated database that contains $9.75$ hours of audio recordings for training. Such an amount is considered inadequate to exploit the full capability of the latest deep learning techniques. In ICASSP 2017, the Sound and Video Understanding team at Google Research announced the release of AudioSet \cite{audioset2017ontology}, which comprises a large amount of audio samples from YouTube. The total duration of data in the current release of AudioSet exceeds $5000$ hours. Unlike the DCASE2016 ASC database, audio samples in AudioSet are labeled by a large number of sound categories which are organised in a loose hierarchy. While the availability of AudioSet has caught great attention from the research community, there have been few published studies that report referable classification performance on the database. In \cite{googlecnnaudioclass}, a related database named YouTube-100M was used to investigate large-scale audio classification problem. The YouTube-100M dataset contains $100$ million YouTube videos. The experimental results show that with massive amount of training data, a residual network with $50$ layers produces the best performance, in comparison with the MLP, AlexNet, VGG and Inception network \cite{googlecnnaudioclass}.

This paper presents our recent attempt to large-scale audio classification with the newly released AudioSet. To our knowledge, except for the preliminary evaluation briefly mentioned in \cite{googlecnnaudioclass}, there has been no official published result on the complete AudioSet classification task. We apply a variety of commonly used DNN models on the AudioSet task and find that CNN based models generally achieve better performance than MLP and RNN. We further propose to exploit low-dimension feature representation of audio segments, so as to achieve significant reduction of CNN model complexity. It is shown that the number of model parameters could be reduced by $22$ times while maintaining comparable performance of classification. In addition, the effectiveness of the proposed methods is validated on the DCASE2016 ASC database.

In Section \ref{sec:dataset}, the AudioSet and the TUT Acoustic Scenes 2016 database are described. The general framework of the classification system and the proposed strategy of model complexity reduction are explained in Section \ref{sec:systemdesign}. Experimental results with different types of neural network models are given in Section \ref{sec:experiments}.

\section{DATASETS FOR AUDIO CLASSIFICATION}
\label{sec:dataset}

%Audio classification experiments presented in this paper are carried out mainly on AudioSet \cite{audioset2017ontology}. The proposed method is also evaluated on the TUT Acoustic Scenes 2016 database \cite{mesaros2016tut}, in order to be benchmarked against the performance of the latest systems.

\subsection{AudioSet}
\label{ssec:audiosetdatabase}

The AudioSet is a large-scale collection of human-annotated audio segments from YouTube \cite{audioset2017ontology}. It is provided as text (csv) files that contain the following attributes of each audio sample: YouTube Video ID, start time and end time of the audio clip, and sound category labels. Each audio clip is $10$ second long. The sound labels were obtained through a human-annotation process, in which human raters were asked to confirm the presence of a set of hypothesized sound categories. Both audio and video components were presented to the raters. The hypothesized sound categories were generated from multiple sources, including a video-labeling system and various meta-data information. There may be multiple sound categories co-existing in an audio clip. There are totally $527$ sound categories being used in AudioSet. These categories are arranged following a loose hierarchy. For example, ``speech'' and ``male speech'' are treated as two categories at different hierarchical levels. However, this kind of hierarchy is not taken into account in the audio classification experiments.

The entire AudioSet contains about $2$ million audio samples, which correspond to more than $5,000$ hours of data. The audio samples are divided into the balanced training set, the unbalanced training set and the evaluation set. In this study, only the balanced training set and evaluation set are used. A number of audio samples are excluded for various reasons, e.g., deleted YouTube links, duration shorter than $10$ seconds. As a result, the number of audio samples used for training and evaluation are $20,175$ and $18,396$ respectively. A validation set is created by randomly selecting $10\%$ of the training data.

\subsection{TUT Acoustic Scenes 2016 database}
\label{ssec:tutdatabase}

The TUT Acoustic Scenes 2016 database \cite{mesaros2016tut} was used in the DCASE2016 challenge. There are $15$ defined acoustic scenes, covering various indoor and outdoor environments. The development dataset contains $1170$ audio samples and the evaluation dataset contains $390$ samples. The number of samples representing different scene classes are the same. Each audio sample is $30$ second long and said to be from one and only one of the $15$ scenes. The total duration of recordings is $13$ hours.

\section{SYSTEM DESIGN}
\label{sec:systemdesign}

\subsection{General System Framework}
\label{ssec:generalframework}

\begin{figure}[h]
  \centering
  \includegraphics[width=0.8\linewidth,keepaspectratio]{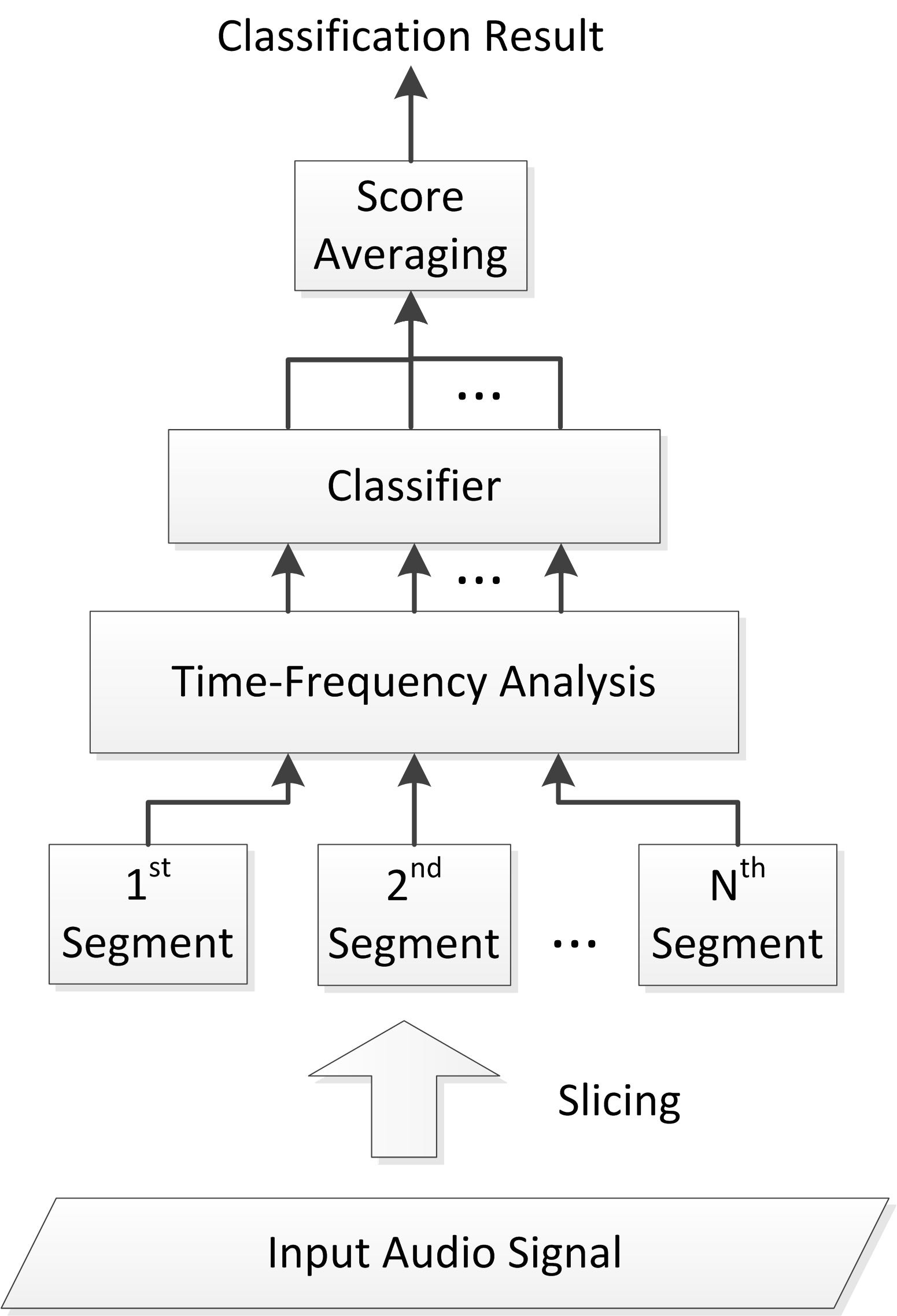}
  \caption{The general framework for a segment-based audio classification system.}
  \label{generalframework}
\end{figure}

Figure \ref{generalframework} shows the general framework of a segment-based audio classification system. The typical length of a segment is 1 second. A segment can be divided into short-time frames (typically $25$ ms long) which are used for spectral analysis. The segment-based system can make better use of the temporal information of audio signals than the frame-based system (e.g., the baseline GMM model in DCASE2016 ASC task \cite{Heittola2016}). In Figure \ref{generalframework}, the input audio signal (e.g., $10$-second audio clip in AudioSet) is divided into non-overlapping segments. The sound category labels of a segment are inherited from those of the input audio signal. For each segment, a time-frequency representation is derived for classification purpose. The time-frequency features of each segment are fed into a classifier to obtain the classification scores. The sample-level classification score is calculated by averaging the segment-level scores.

Commonly used time-frequency representations for audio classifications are derived from the short-time Fourier transforms. Examples include log-mel filterbank features, Constant-Q transform (CQT) \cite{calCQT}, and MFCC. Based on our preliminary experiments, for DNN-based systems, the log-mel features give the best performance among these feature types, and thus are used in our experiments.

For the classifier in Figure \ref{generalframework}, our main focus of experiments is the DNN models. There has been growing interest in extracting the embedding features from a well-trained DNN classifier in audio classification area. For example, Rakib et al. \cite{cnnsvplda} uses the trained CNN model to extract its embedding feature, which is fed into PLDA to improve classification performance. In \cite{Mun2017}, the use of embedding features from a trained DNN classifier also serve as a critical component in its proposed method. In this paper, several ways to obtain low-dimensional embedding feature are studied in Section \ref{ssec:lowdimfeature}.

\subsection{Strategy for Reducing Model Complexity}
\label{ssec:strategies}

\subsubsection{Use of Bottleneck Layer}
\label{sssec:bottleneck}

%[gives background, application, intention of using bottleneck]
%[show related work using bottleneck layers for different purposes, and then state your purpose.]

Bottleneck layer has been applied in speech recognition area to extract embedding features. The extracted feature is called bottleneck feature and is generally better than the hand-crafted feature \cite{bottleneck}. Recently, bottleneck layers were investigated in large-scale audio classification by Shawn Hershey et al. \cite{googlecnnaudioclass}. The introduction of bottleneck layer leads to faster training, while maintaining comparable classification performance.

\begin{figure}[h]
  \centering
  \includegraphics[width=\linewidth,keepaspectratio]{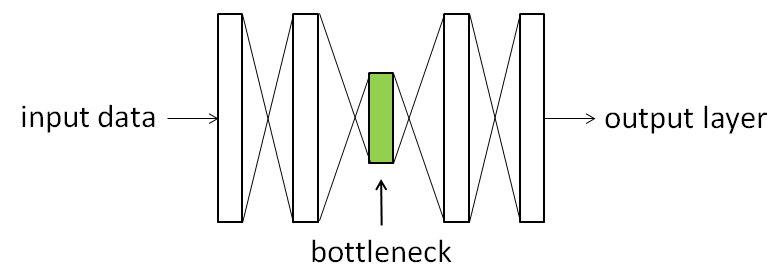}
  \caption{An illustration of bottleneck layer in a multi-layer perceptron (MLP). The bottleneck layer has smaller size than its adjacent layers.}
  \label{illbneck}
\end{figure}

A bottleneck layer typically lies in between two (hidden) layers in a fully-connected neural network. It is a middle layer designed to have a relatively small number of neurons as compared to other hidden layers, and therefore called ``bottleneck''. By constructing a bottleneck layer, a low-dimensional feature representation of input data can be generated. Figure \ref{illbneck} shows an example of MLP with a bottleneck layer.

In this study, we make use of the bottleneck layers to achieve reduction of model complexity, and through which low-dimensional embedding feature extractor is constructed. Different sizes of bottleneck layer are experimented to reveal the trade-off between model performance and complexity.

\subsubsection{Global Average Pooling}
\label{sssec:gloablavgpooling}

A CNN-based classification model is typically composed of convolution layer(s) and fully connected (FC) layer(s). The common way of making connection between a convolution layer and a FC layer is by flattening (vectorizing) the feature maps of the convolutional layer and using the flattened features as the input of the FC layer. Since the flattened features have a very large dimension, the number of required model parameters would be excessive. Moreover, it may increase the chance of over-fitting of the FC layers.

%[Give a more general background about GAP, e.g., its initial motivation, its most known success (image), with citations].

\begin{figure}[h]
  \centering
  \includegraphics[width=\linewidth,keepaspectratio]{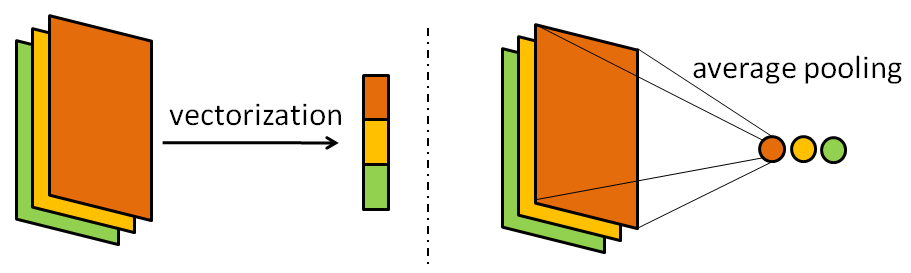}
  \caption{Left: conventional way of linking up convolutional layer and fully connected layer; right: global average pooling.}
  \label{global_avg_pool}
\end{figure}

In \cite{netinnet}, global average pooling strategy is proposed to solve the problem of over-fitting of FC layers, and its effectiveness of being a regularizer has been verified. It is an average pooling operation applied on each feature map obtained from the last convolutional layer, with the size of pooling window equal to the size of feature map. This pooling result is used as the input of FC layer(s) for classification. Figure \ref{global_avg_pool} illustrates the conventional way and global average pooling to transform 2D feature maps into 1D feature vector.

In this study, we emphasize global average pooling for its efficacy of reducing CNN model complexity, while preserving the performance of classification.

\section{EXPERIMENTS}
\label{sec:experiments}

\subsection{EXPERIMENTAL SETUP}
\label{ssec:experimentsetup}

\subsubsection{Data Preprocessing}
\label{sssec:preprocessdata}

Each audio sample in the dataset is divided into non-overlapping $1$-second segments. The sound category labels assigned to each segment are exactly the same as the source audio sample. Short-Time Fourier Transform (STFT) is applied on the audio segments with a window length of $25$ ms, hop length of $10$ ms and FFT length of $2048$. Subsequently $64$-dimensional log-mel filterbank features are derived from each short-time frame, and the frame-level features are put together to form a time-frequency matrix representation of the segment. Dimension-wise normalization of the log-mel features is performed using the means and variances calculated from all audio samples in the training set.

\subsubsection{Performance Metric}
\label{sssec:evaluationmetric}

The Area Under Receiver Operating Characteristic curve, abbreviated as AUC \cite{rocaucdef}, is used as our performance metric in the experiments with the AudioSet. In the context of binary classification, AUC can be viewed as the probability that the classifier ranks a randomly chosen positive sample higher than a negative one \cite{aucproperty}. For a classification model only makes random guesses, the AUC value is $0.5$. A perfect classification model gives the AUC value of $1.0$. AUC is found to be insensitive to the distribution of positive and negative samples, as compared to other evaluation metrics like precision, accuracy, $F1$ score and mAP.

For a multi-class problem, the overall measure of AUC is obtained as the weighted average of the AUC values for individual classes. The weight for a specific class is proportional to its prevalence in the dataset.
%For example, if ``speech'' class occupies $21\%$ of the total labels, its weight will be $0.21$.

\subsubsection{Model Training and Parameter Setting}
\label{sssec:trainstrategy}

%Audio classification experiments are carried out with three major categories of neural networks, namely MLP, RNN and CNN, using the PyTorch toolbox for deep learning (http://pytorch.org/).

The experimented models in this study are implemented using the deep learning toolbox PyTorch (http://pytorch.org/). The key parameters used for training are empirically determined. The initial learning rate and mini-batch size are set to $0.0001$ and $60$. Model training is done by minimizing the cross-entropy loss with the Adam optimizer (${\beta _1} = 0.9$ and ${\beta _2} = 0.999$) and learning rate decay strategy. For MLP and CNN models, dropout (with dropout probability $= 0.5$) and weight decay (coefficient $=0.0015$) are applied for regularization purpose. The sigmoid function is used in the output layer for all models, considering that an audio sample may have multiple sound labels.

\subsection{Model Comparison}
\label{ssec:comparemodels}

Table \ref{tb-model-compare} shows the experimental results with six different models (or different model configurations). The MLP model has $3$ hidden layers with $1000$ neurons per layer. For the MLP, batch normalization and the ReLU activation function are applied. The LSTM (Long-Short Term Memory) model contains $3$ LSTM layers, each having $2048$ units. GRU refers to Gated Recurrent Unit \cite{gru}. B-GRU-ATT refers to bi-directional GRU with its output weighted by attention network \cite{Yang2016HierarchicalAN} whose context vector size is $1024$. It has $2$ GRU layers, each with $2048$ units. To our knowledge, performance of recurrent models has not been reported for AudioSet yet.

CNN models have been investigated for large-scale audio classification on YouTube-100M database in \cite{googlecnnaudioclass}. In this study, the CNN models being experimented are the AlexNet and Residual Network with 50 layers (ResNet-50) \cite{deepresnet}. The AlexNet used in our experiments is similar to the AlexNet described in \cite{newalexnet}, which was designed for image classification with $224 \times 224 \times 3$ input. We make a change of its first convolutional layer to have a kernel size of $11 \times 7$ and stride of $2 \times 1$, so as to obtain a similar size of feature map at the first convolutional layer. For the FC layers, the size is set to $3982$. By AlexNet(BN), we mean that a batch normalization layer is added after each layer (convolutional and FC).

For the ResNet-$50$ model, we follow the same setting as in \cite{googlecnnaudioclass}, by changing the stride to $1$ in the first convolution layer. As a result, the window length of its global average pooling layer is set to $7\times 4$, to match the change of feature map size. The model sizes given in the table refer to the number of model parameters in the respective models.

The overall AUC is calculated over $527$ audio classes (see Section \ref{sssec:evaluationmetric}). It can be seen that the AlexNet with batch normalization performs the best among all tested models. It even outperforms the $50$-layer deep residual network, which was reported to have the highest performance among CNN models for large-scale audio classification \cite{googlecnnaudioclass}.

\begin{table}[]
\centering
\caption{Classification performance of six DNN models tested on AudioSet evaluation set, trained with the balanced training set. The letter ``M'' in ``Model Size'' column stands for million.}
\label{tb-model-compare}
\begin{tabular}{|l|l|l|l|}
\hline
Model       & Structure & Model Size       & AUC   \\ \hline
MLP         & $3\times1000$    & $9.48$M            & $0.845$ \\ \hline
LSTM        & $3\times2048$    & $85.54$M           & $0.866$ \\ \hline
B-GRU-ATT   & $2\times2048$    & $107.85$M          & $0.870$ \\ \hline
AlexNet     & -                & $56.09$M           & $0.895$ \\ \hline
\textbf{AlexNet(BN)} & \textbf{-}      & \textbf{$\mathbf{56.11}$M}     & \textbf{$\mathbf{0.927}$} \\ \hline
ResNet-50   & -                & $24.58$M           & $0.914$ \\ \hline
\end{tabular}
\end{table}

\subsection{Reducing Model Complexity}
\label{ssec:lowdimfeature}

While the AlexNet(BN) model has been shown to have the best performance among different DNN models, its model complexity is relatively large and thus undesirable for practical use. As described in Section \ref{ssec:strategies}, using a bottleneck layer and performing global average pooling are effective techniques of reducing the number of model parameters. We experiment with different arrangements of the FC layers and the bottleneck layer in the AlexNet(BN) model. The results are compared as in Table \ref{dimreducedmodel}. ``Bneck-Final-64'' refers to that a $64$-dimension bottleneck layer is inserted between the output layer and the last FC layer, while ``Bneck-Mid-64'' means that the $64$-dimension bottleneck layer is inserted between the two FC layers. For the ``FC-64''  configuration, the size of both FC layers is reduced to $64$. Three different sizes of embedding features are tested: $64$, $256$ and $1024$. Lastly, ``Global-avg-pool'' means that a global average pooling layer is used to replace the two FC layers. The resulting feature dimension after pooling is $256$, which is equal to the number of feature maps in the last convolution layer.

Generally, a larger size of bottleneck layer or FC layers lead to better classification performance. Reducing the size of existing FC layers without having an additional bottleneck layer would cause noticeable degradation of performance, despite the significantly reduced model complexity. With the same size of bottleneck layer, it is more beneficial to have the bottleneck inserted between the two FC layers (i.e., the ``Bneck-Mid'' configurations). ``Bneck-Mid-$1024$'' could attain the same AUC as the original AlexNet(BN), with about $14\%$ less model parameters. By applying global average pooling strategy, the model complexity is reduced to $2.59$M. which is about $1/22$ of the original AlexNet(BN), $1/9$ of the ResNet-$50$ model, and $1/4$ of the MLP model. Its performance is comparable to the ResNet-$50$ model, and slightly worse than the AlexNet(BN).

\begin{table}[]
\centering
\caption{Performance of 4 types of strategies for reducing model complexity. All strategies are applied on the same AlexNet(BN) model as described in Section \ref{ssec:comparemodels}.}
\label{dimreducedmodel}
\begin{tabular}{|l|l|l|}
\hline
Strategy           & Model Size           & AUC   \\ \hline
None               & $56.11$M               & $0.927$ \\ \hline
Bneck-Final-$64$   & $54.30$M               & $0.889$ \\ \hline
Bneck-Final-$256$  & $55.17$M               & $0.917$ \\ \hline
Bneck-Final-$1024$ & $58.63$M               & $0.925$ \\ \hline
Bneck-Mid-$64$     & $40.77$M               & $0.915$ \\ \hline
Bneck-Mid-$256$    & $42.29$M               & $0.924$ \\ \hline
Bneck-Mid-$1024$   & $48.41$M               & $0.927$ \\ \hline
FC-$64$            & $3.07$M                & $0.841$ \\ \hline
FC-$256$           & $4.95$M                & $0.905$ \\ \hline
FC-$1024$          & $13.22$M               & $0.924$ \\ \hline
Global-avg-pool    & $\mathbf{2.59}$M       & $0.916$ \\ \hline
\end{tabular}
\end{table}

\subsection{Acoustic Scene Classification in DCASE2016}
\label{ssec:ascresults}

The proposed models are also evaluated with the TUT Acoustic Scenes 2016 database. Due to the different nature of the scene classification task, softmax function is used at the output layers of the neural networks. The other settings of training are the same as stated in Section \ref{sssec:trainstrategy}. Among the $1170$ audio samples in the training set, $170$ samples are randomly selected to be the validation data. The AlexNet(BN) model attains a classification accuracy of $87.4\%$ on the evaluation set, while a $3$-layer MLP with $1000$ neurons per layer has an accuracy of $78.2\%$, and a well-tuned LSTM model has an accuracy of $82.8\%$. By applying the strategy of global average pooling, the size-reduced AlexNet(BN) has an accuracy of $85.9\%$. This further confirms the effectiveness of CNN and global average pooling strategy, though the ASC may not be viewed as a large-scale task as compared to the AudioSet.

\section{CONCLUSION}
\label{sec:conclusion}

The AudioSet database provides useful resources to enable and advance research on large-scale audio classification. This paper presents one of the earliest batches of experimental results on this database using the latest neural network models. It has been shown that CNN models are more effective than MLP and RNN. The model complexity of the best-performing CNN can be significantly reduced by introducing a bottleneck layer at the fully-connected layers and by applying global average pooling. It must be noted that only a small portion of AudioSet has been used in the present study, though this small portion already contains $20$ times more audio samples than the existing DCASE2016 ASC database.

%\vfill
%\pagebreak


% Start a new page which only contains references
% \clearpage

\bibliographystyle{IEEEbib}
\bibliography{refs}

\end{document}